%% file: main.tex
\documentclass{article}
\usepackage{arxiv}
\usepackage[utf8]{inputenc} 
\usepackage[T1]{fontenc}    
\usepackage{hyperref}       
\usepackage{url}            
\usepackage{booktabs}       
\usepackage{amsfonts}       
\usepackage{nicefrac}       
\usepackage{microtype}      
\usepackage{fancyhdr}       
\usepackage{graphicx}       
\usepackage{float} 
\usepackage[T1]{fontenc}
\usepackage{dcolumn}
\usepackage{bm}
\usepackage{siunitx}
\usepackage{mathptmx}
\usepackage{etoolbox}
\usepackage{amsmath}
\usepackage{amssymb}
\usepackage{tabularx}
\usepackage{soul}
\setstcolor{red}
\DeclareSIUnit\angstrom{\text {Å}}
\usepackage{multirow}
\usepackage[style=chem-acs]{biblatex}
\title{Accelerated Bayesian Inference for Molecular Simulations using Local Gaussian Process Surrogate Models
}

\author{
  B. L. Shanks, H. W. Sullivan, A. R. Shazed, M. P. Hoepfner \\
  Department of Chemical Engineering \\
  University of Utah \\
  Salt Lake City, UT\\
  \texttt{B. L. Shanks} {brennon.shanks@chemeng.utah.edu} \\
  \texttt{M. P. Hoepfner} {michael.hoepfner@utah.edu}
}

\begin{document}

\maketitle

\begin{abstract}

While Bayesian inference is the gold standard for uncertainty quantification and propagation, its use within physical chemistry encounters formidable computational barriers. These bottlenecks are magnified for modeling data with many independent variables, such as X-ray/neutron scattering patterns and electromagnetic spectra. To address this challenge, we employ local Gaussian process (LGP) surrogate models to accelerate Bayesian optimization over these complex thermophysical properties. The time-complexity of the LGPs scales linearly in the number of independent variables, in stark contrast to the computationally expensive cubic scaling of conventional Gaussian processes. To illustrate the method, we trained a LGP surrogate model on the radial distribution function of liquid neon and observed a 1,760,000-fold speed-up compared to molecular dynamics simulation, beating a conventional GP by three orders-of-magnitude. We conclude that LGPs are robust and efficient surrogate models, poised to expand the application of Bayesian inference in molecular simulations to a broad spectrum of experimental data.

\end{abstract}

\section{Introduction}

Molecular simulations are able to estimate a broad array of complex experimental observables, including scattering patterns from neutron and X-ray sources and spectra from near-infrared \cite{czarnecki_advances_2015}, terahertz \cite{schmuttenmaer_exploring_2004}, sum frequency generation \cite{nihonyanagi_ultrafast_2017, hosseinpour_structure_2020}, and nuclear magnetic resonance \cite{mishkovsky_principles_2009}. Recent interest in these experiments to study hydrogen bonding networks of water at interfaces \cite{roget_water_2022, li_hydrogen_2022}, electrolyte solutions \cite{wang_hydrogen-bond_2022}, and biological systems \cite{meng_modeling_2023} has motivated the continued advancement of simulations to calculate these properties from first-principles \cite{bally_quantum-chemical_2011, thomas_computing_2013, gastegger_machine_2017}. However, the ability to estimate these complex properties comes with a high computational cost. This barrier greatly limits our ability to quantify how experimental, model, and parametric uncertainty impact molecular simulation predictions, making it difficult to know whether a model is an appropriate representation of nature or if it is simply over-fitting to a given training set. Therefore, what is needed is a computationally efficient and rigorous uncertainty quantification/propagation (UQ/P) method to link molecular models to large and complex experimental datasets.

Bayesian methods are the gold standard for these aims \cite{psaros_uncertainty_2023}, with examples spanning from neutrino and dark matter detection \cite{eller_flexible_2023}, materials discovery and characterization \cite{todorovic_bayesian_2019, zuo_accelerating_2021, fang_exploring_2023, sharmapriyadarshini_pal_2024}, quantum dynamics \cite{vkrems_bayesian_2019, deng_bayesian_2020}, to molecular simulation \cite{frederiksen_bayesian_2004, cooke_statistical_2008, cailliez_statistical_2011, farrell_bayesian_2015, wu_hierarchical_2016, patrone_uncertainty_2016, messerly_uncertainty_2017, dutta_bayesian_2018, wen_uncertainty_2020, bisbo_efficient_2020, xie_uncertainty-aware_2023}. The Bayesian probabilistic framework is a rigorous, systematic approach to quantify probability distribution functions on model parameters and credibility intervals on model predictions, enabling robust and reliable parameter optimization and model selection \cite{gelman_bayesian_1995, shahriari_taking_2016}. Interest in Bayesian methods and uncertainty quantification for molecular simulation has surged \cite{musil_fast_2019,cailliez_bayesian_2020, vandermause_--fly_2020, kofinger_empirical_2021, vandermause_active_2022, blasius_uncertainty_2022} due to its flexible and reliable estimation of uncertainty, ability to identify weaknesses or missing physics in molecular models, and systematically quantify the credibility of simulation predictions. Additionally, standard inverse methods including relative entropy minimization, iterative Boltzmann inversion, and force matching have been shown to be approximations to a more general Bayesian field theory \cite{lemm_bayesian_2003}.

The biggest problem plaguing Bayesian inference is its massive computational cost. The two major pinch points are (1) sampling in high-dimensional spaces, commonly known as the "curse of dimensionality", and (2) the large number of model evaluations required to get accurate uncertainty estimates. In computational chemistry, these bottlenecks are magnified since these models are typically expensive. Therefore, rigorous and accurate uncertainty estimation is challenging, or even impossible, without accelerating the simulation prediction time. One way to achieve this speed-up is by approximating simulation outputs with an inexpensive machine learning model. These so-called surrogate models have been developed from neural networks \cite{wen_uncertainty_2020, li_rapid_2023}, polynomial chaos expansions \cite{ghanem_stochastic_2003, jacobson_how_2014}, configuration-sampling-based methods \cite{messerly_configuration-sampling-based_2018} and Gaussian processes \cite{rasmussen_gaussian_2006, nguyen-tuong_model_2009, burn_creating_2020}. 

Gaussian processes (GPs) are a compelling choice as surrogate models thanks to several distinct advantages. GPs are non-parametric, kernel-based function approximators that can interpolate function values in high-dimensional input spaces. GPs with an appropriately selected kernel also have analytical derivatives and Fourier transforms, making them well-suited for physical quantities such as potential energy surfaces \cite{dai_interpolation_2020, yang_local_2023}. Additionally, kernels can encode physics-informed prior knowledge, alleviating the "black box" nature inherent to many machine learning algorithms. In fact, a comparison of various nonlinear regressors for molecular representations of ground-state electronic properties in organic molecules demonstrated that kernel regressors drastically outperformed other techniques, including convolutional graph neural networks \cite{faber_prediction_2017}. 

Perhaps the most widely adopted application of GP surrogate models in computational chemistry is for model optimization. In the last decade, GP surrogates of simple thermophysical properties including density, heat of vaporization, enthalpy, diffusivity and pressure have been used for force field design 
\cite{angelikopoulos_bayesian_2012,cailliez_calibration_2014, kulakova_data_2017,befort_machine_2021, cmadin_using_2023, wang_machine_2023}. However, to our knowledge there are no Bayesian optimization studies that apply GP surrogate models to thermophysical properties with many independent variables, such as structural correlation functions or electromagnetic spectra. In this work, independent variables (IVs) are defined as the fixed quantities over which a measurement is made (\textit{e.g.} frequencies along a spectrum or radial positions along a radial distribution function) and the outcomes of those measurements are referred to as quantities-of-interest (QoIs). 

Measurements of complex QoIs with many IVs are often available or easily obtained, yet are rarely included as observations in Bayesian optimization of molecular models. One reason why this may be the case is that previous literature has not outlined accurate and robust approaches to design Gaussian process surrogates for such data. For example, Angelikopoulous and coworkers did not use GP surrogate models for their Bayesian analysis on the radial distribution function (RDF) of liquid Ar \cite{angelikopoulos_bayesian_2012}, despite the fact that doing so would significantly reduce computation time. It is likely that GPs have not been previously used for complex QoIs due to high training and evaluation costs. Specifically, GPs have a cubic time-complexity in the number of IVs, which quickly becomes prohibitively expensive as experimental measurements obtain higher ranges and resolutions. 

Local Gaussian processes (LGPs) are an emerging class of accelerated GP methods that are well-equipped to handle large sets of experimental data. These so-called "greedy" Gaussian process approximations are constructed by separating a GP into a subset of GPs trained at distinct locations in the input space \cite{nguyen-tuong_model_2009, das_block-gp_2010, park_patchwork_2018, terry_splitting_2021}. Computation on the LGP subset scales linearly with the number of IVs, is trivially parallelizable, and easily implemented in high-performance computing (HPC) architectures \cite{gramacy_local_2015, broad_parallel_2023}. State-of-the-art LGP models have been used to design Gaussian approximation potentials (GAPs) \cite{deringer_gaussian_2021}, a type of machine learning potential used to study atomic \cite{caro_growth_2018, deringer_realistic_2018, lderinger_towards_2018} and electron structures \cite{cheng_evidence_2020, deringer_gaussian_2021}, as well as nuclear magnetic resonance chemical shifts \cite{paruzzo_chemical_2018} with uncertainty quantification \cite{musil_fast_2019}. However, to our knowledge LGPs have not been applied as surrogate models for UQ/P on complex experimental data in computational chemistry.

In this study, we detail a simple and effective surrogate modeling approach for complex experimental observables common in physical chemistry. LGPs unlock the capability for existing Bayesian optimization schemes to incorporate complex data efficiently and accurately at a previously inaccessible computational scale. The key feature of the LGP surrogate model is the reduction in time-complexity with respect to the number of QoIs from cubic to linear, resulting in orders-of-magnitude speed-ups to evaluate complex observable surrogate models and perform posterior estimation. The computational speed-up results from reducing the dimensionality of matrix operations and therefore enables Bayesian UQ/P on experimental data with many IVs. For illustration, consider that a typical Fourier transformed infrared spectroscopy (FT-IR) measurement may contain data between 4000-400 cm$^{-1}$ at a resolution of 2 cm$^{-1}$, giving a total number of QoIs around $\eta = 1800$. According to the time-complexity scaling in $\eta$, a LGP is estimated to accelerate this computation compared to a standard GP by approximately 3,240,000x. Source code and a tutorial on building LGP surrogate models is provided on GitHub.

To demonstrate the method, we trained a LGP surrogate model on the RDF of the ($\lambda$-6) Mie fluid and performed Bayesian optimization to fit the parameters of the Mie fluid model to a neutron scattering derived RDF for liquid neon (Ne). The LGP was found to accelerate the $\eta=73$ independent variable surrogate model calculation approximately 1,760,000x faster than molecular dynamics (MD) and 2100x faster than a conventional GP with accuracy comparable to the uncertainty in the reported experimental data. Bayesian posterior distributions were then calculated with Markov chain Monte Carlo (MCMC) and used to draw conclusions on model behavior, uncertainty, and adequacy. Surprisingly, we find evidence that Bayesian inference conditioned on the radial distribution function significantly constrains the ($\lambda$-6) Mie parameter space, highlighting opportunities to improve force field optimization and design based on neutron scattering experiments.      

\section{Computational Methods}

In the following sections, an outline of standard approaches for Bayesian inference and surrogate modeling with Gaussian processes is presented. Then, we describe the local Gaussian process approximation and highlight key differences in their implementation and computational scaling.  

\subsection{Bayesian Inference}

Bayes' law, derived from the definition of conditional probability, is a formal statement of revising one's prior beliefs based on new observations. Bayes' theorem for a given model, set of model input parameters, $\boldsymbol{\theta}$, and set of experimental QoIs, $\mathbf{y}$, is expressed as,  

\begin{equation}\label{eq:bayes}
    p(\boldsymbol{\theta}|\mathbf{y}) \propto p(\mathbf{y}|\boldsymbol{\theta}) p(\boldsymbol{\theta})
\end{equation}

\noindent where $p(\boldsymbol{\theta})$ is the 'prior' probability distribution over the model parameters, $p(\mathbf{y}|\boldsymbol{\theta})$ is the 'likelihood' of observing $\mathbf{y}$ given parameters $\boldsymbol{\theta}$, and $p(\boldsymbol{\theta}|\mathbf{y})$ is the 'posterior' probability that the underlying parameter $\boldsymbol{\theta}$ models or explains the observation $\mathbf{y}$. Equality holds in eq \eqref{eq:bayes} if the right-hand-side is normalized by the 'marginal likelihood', $p(\mathbf{y})$, but including this term explicitly is unnecessary since the posterior probability distribution can be normalized \textit{post hoc}. In molecular simulations, $\boldsymbol{\theta}$ is the set of unknown parameters in the selected model, usually the force field parameters in the Hamiltonian, to the experimental QoI that the simulation estimates. The observations, $\mathbf{y}$, can be any QoI or combination of QoIs (\textit{e.g.} RDFs, spectra, densities, diffusivities, etc). This construction, known as the standard Bayesian scheme, is generalizable to any physical model and its corresponding parameters including density functional theory (DFT), \textit{ab initio} molecular dynamics (AIMD), and path integral molecular dynamics (PIMD).

Calculating the posterior distribution then just requires prescription of prior distributions on the model input parameters and evaluation of the likelihood function. In this work, Gaussian distributions are used for both the prior and likelihood functions, which is a standard choice according to the central limit theorem. The Gaussian likelihood has the form,

\begin{equation}\label{eq:likelihood}
    p(\mathbf{y}|\boldsymbol{\theta}) = \bigg(\frac{1}{\sqrt{2 \pi}\sigma_n}\bigg)^\eta \exp\bigg[-\frac{1}{2\sigma_n^2}\sum_{i=1}^{\eta}\ [\mathbf{y}_{\boldsymbol{\theta}} - \mathbf{y}]^2\bigg]
\end{equation}

\noindent where $\eta$ is the number of observables in $\mathbf{y}$, $\mathbf{y}_{\boldsymbol{\theta}}$ is the model predicted observables at model input $\boldsymbol{\theta}$, and $\sigma_n$ is a nuisance parameter describing the unknown variance of the Gaussian likelihood. Cailliez and coworkers choose the nuisance parameter as the sum of simulation and experiment variances ($\sigma_n^2 \approx \sigma_{sim}^2 + \sigma_{exp}^2$) \cite{cailliez_calibration_2014}; however, if these variances are unknown or one wishes to explore the distribution of variances, the nuisance parameter can be inferred via the Bayesian inference. Hence, the resulting posterior distribution on the nuisance parameter includes the unknown uncertainty arising due to the sum of the model and the experimental variances. In this work, the nuisance parameter is treated as an unknown to be inferred along with the explicit model parameters. Note that in some cases a different likelihood function may be more appropriate based on physics-informed prior knowledge of the distribution of the observable of interest (\textit{e.g.} the multinomial likelihood in relative entropy minimization between canonical ensembles \cite{shell_relative_2008}).

The computationally expensive part of calculating eq \ref{eq:likelihood} is determining $\mathbf{y}_{\boldsymbol{\theta}}$ at a sufficient number of points in the parameter space. Generally, this can be achieved by calculating $\mathbf{y}_{\boldsymbol{\theta}}$ at dense, equally spaced points in the parameter space of interest (grid method), sampling the parameter space with Markov chain Monte Carlo (MCMC) to estimate the posterior with a histogram (approximate sampling method), or assuming that the posterior distribution has a specific functional form (\textit{i.e.} Laplace approximation). Regardless of the selected method, each of these posterior distribution characterization techniques require a prohibitive number of molecular simulations to adequately sample the parameter space (often on the order of $10^5-10^6$), which is infeasible for even modest sized molecular systems.

\subsection{Gaussian Process Surrogate Models}

Gaussian processes accelerate the Bayesian likelihood evaluation by approximating $\mathbf{y}_{\boldsymbol{\theta}}$ with an inexpensive matrix calculation. A Gaussian process is a stochastic process such that every finite set of random variables (position, time, etc) has a multivariate normal distribution \cite{rasmussen_gaussian_2006}. The joint distribution over all random variables in the system therefore defines a functional probability distribution. The expectation of this distribution maps a set of model parameters, $\boldsymbol{\theta}^*$, and IVs, $\mathbf{r}$, to the most probable QoI given the model parameters, $S(\mathbf{r} | \boldsymbol{\theta}^*)$, such that,

\begin{equation}
    \mathbb{E}[GP] : \boldsymbol{\theta}^* \times \mathbf{r} \mapsto S(\mathbf{r} | \boldsymbol{\theta}^*)
\end{equation}

\noindent where the expectation operator is written in terms of a kernel matrix, $\mathbf{K}$, training set parameter matrix, $\mathbf{\hat{X}}$, and training set output matrix, $\mathbf{\hat{Y}}$, according to the equation,

\begin{equation}\label{eq:surrogate}
    \mathbb{E}[\textit{GP}(\boldsymbol{\theta}^*, \mathbf{r})] = \mathbf{K}_{(\boldsymbol{\theta}^*, \mathbf{r}),\mathbf{\hat{X}}} [\mathbf{K}_{\mathbf{\hat{X}}, \mathbf{\hat{X}}} + \sigma_{noise}^2 \mathbf{I}]^{-1} \mathbf{\hat{Y}}
\end{equation}

\noindent where $\sigma_{noise}^2$ is the variance due to noise and $\mathbf{I}$ is the identity matrix. Note that in general the IVs, $\mathbf{r}$, can be multidimensional. As an example, consider the case a GP maps a set of force field parameters to the angular RDF of a liquid. We now have a 2-dimensional space of IVs since the angular RDF gives the atomic density along the radial and angular dimensions. In the following mathematical development, it is assumed that the QoI is 1-dimensional for sake of convenience and note that extending the method to higher-dimensional observables just requires redefining the IVs in accordance with eq \eqref{eq:surrogate}.

The kernel matrix, $\mathbf{K}$, quantifies the relatedness between input parameters and can be selected based on prior knowledge of the physical system. A standard kernel for physics-based applications is the squared-exponential (or radial basis function) since the resulting GP is infinitely differentiable, smooth, continuous, and has an analytical Fourier transform \cite{ambrogioni_integral_2018}. The squared-exponential kernel function between input points $(\boldsymbol{\theta}_m,
r_m)$ and $(\boldsymbol{\theta}_n, r_n)$ is given by,

\begin{equation}\label{eq:kernel}
    K_{mn} = \alpha^2 \exp\bigg(-\frac{(r_{m} - r_{n})^2}{2\ell_{r}^2} - \sum_{o=1}^{\text{dim}(\boldsymbol{\theta})}\frac{(\theta_{o,m} - \theta_{o,n})^2}{2\ell_{\theta_o}^2} \bigg)
\end{equation}

\noindent where $o$ indexes over dim($\boldsymbol{\theta}$) and the hyperparameters $\alpha^2$ and $\ell_A$ are the kernel variance and correlation length scale of parameter $A$, respectively. Hyperparameter optimization can be performed by log marginal likelihood maximization, $k$-fold cross validation \cite{rasmussen_gaussian_2006} or marginalization with an integrated acquisition function \cite{snoek_practical_2012}, but can be computationally expensive and is usually avoided if accurate estimates of the hyperparameters can be made from prior knowledge of the chemical system.

To train a standard GP surrogate model, $N$ training samples are generated in the input parameter space and a molecular simulation is performed for each training set sample to calculate $N$ predictions over the number of target QoIs, $\eta$. The training set, $\mathbf{\hat{X}}$, is then a ($N\eta$ $\times$ dim($\boldsymbol{\theta}$) + 1) matrix of the following form,

\begin{equation}
    \mathbf{\hat{X}} = 
        \begin{bmatrix}
        \theta_{1,1} & \theta_{2,1} & \hdots & r_1\\
        \theta_{1,1} &  \theta_{2,1} & \hdots & r_2\\
        \vdots & \vdots & \vdots & \vdots\\
        \theta_{1,1} & \theta_{2,1} & \hdots & r_\eta\\
        \theta_{1,2} & \theta_{2,2} & \hdots & r_1\\
        \vdots & \vdots & \vdots & \vdots\\
        \theta_{1,N} & \theta_{2,N} & \hdots & r_\eta\\
        \end{bmatrix}
\end{equation}

\noindent where the $\theta_{i,j}$ are the $i^{th}$ model parameter for sample index $j$ and $r_k$ are the IVs of the target QoI. Note that the training sample index, $j = 1,...,N$, is updated in the model parameters only after $\eta$ rows spanning the domain of the observable, giving $N\eta$ total rows. Therefore, the training set matrix represents all possible combinations of the training parameters in the $\boldsymbol{\theta}$ parameter input space.     
The training set observations, $\mathbf{\hat{Y}}$, are a ($N\eta$ $\times$ 1) column vector of the observable outputs from the training set,

\begin{equation}
    \mathbf{\hat{Y}} = [S(\boldsymbol{\theta}_1,r_1), ..., S(\boldsymbol{\theta}_1,r_\eta), S(\boldsymbol{\theta}_2,r_1), ..., S(\boldsymbol{\theta}_N,r_\eta)]^T
\end{equation}

\noindent where $S(\boldsymbol{\theta}_j,r_k) = y(\boldsymbol{\theta}_j,r_k) - \mu_{GP}^{prior}(\boldsymbol{\theta}_j,r_k)$ is the difference between the training set observation of model parameters $\theta_j$ at IV $r_k$ and a GP prior mean function. Of course, the GP prior mean, $\mathbf{\mu}_{GP}^{prior}$, is the same shape as the training set observations matrix,

\begin{equation}
    \mathbf{\mu}_{GP}^{prior} := [\mu(\boldsymbol{\theta}_1,r_1), ..., \mu(\boldsymbol{\theta}_1,r_\eta), \mu(\boldsymbol{\theta}_2,r_1), ..., \mu(\boldsymbol{\theta}_N,r_\eta)]^T
\end{equation}

\noindent where $\mu(\boldsymbol{\theta}_j,r_k)$ is the GP prior mean for parameter set $\boldsymbol{\theta}_j$ at $r_k$. Note that the selection of a prior mean can impact the quality of fit of the GP surrogate model and should reflect physically justified prior knowledge of the physical system. 

Conceptually, since a Gaussian process is a Bayesian model, the prior serves as a current state of knowledge that can encode an initial guess for the QoI before the GP sees any training data. The subtraction of the GP prior mean from the model output effectively shifts the QoI by this pre-specified mean function. Hence, the GP is trained on these mean shifted observations rather than the observations themselves. Although shifting the data by another function seems like it shouldn't change the ability of the GP to estimate the QoI, it actually can have an important impact on the stochastic properties of the data as a function of the IVs. By construction, GPs are stationary, meaning that the means, variances, and covariances are assumed to be equal along all QoI. But for complex data, this is often not the case. For example, it is known that the RDF is zero for small $r$ values and has asymptotic tailing behavior to unity at long-range. The GP prior mean effectively shifts this non-stationary data and makes it behave as if it were stationary by removing any $r$ dependencies.

The expectation of the GP for a new set of parameters, $S^*(\mathbf{r}|\boldsymbol{\theta}^*)$, is then a ($\eta$ x 1) column vector calculated with eq \eqref{eq:surrogate},

\begin{equation}
    S^*(\mathbf{r}|\boldsymbol{\theta}^*) = [S^*(r_1|\boldsymbol{\theta}^*), ..., S^*(r_\eta|\boldsymbol{\theta}^*)]^T
\end{equation}

\noindent where $S^*(\mathbf{r}|\boldsymbol{\theta}^*)$ is the most probable difference function between the model and GP prior mean. Hence, to obtain a comparison to the experimental QoI you simply add the GP prior mean at $\boldsymbol{\theta}^*$, $\mathbf{\mu}_{GP}^{*,prior}(\boldsymbol{\theta}^*, \mathbf{r})$, back to $S^*(\mathbf{r}|\boldsymbol{\theta}^*)$.

The GP expectation calculation is burdened by the inversion of the training-training kernel matrix with $\mathcal{O}(N^3 \eta^3)$ time complexity and the ($\eta$ $\times$ $N\eta$) $\times$ ($N\eta$ $\times$ $N\eta$) $\times$ ($N\eta$ $\times$ 1) matrix product with $\mathcal{O}(N^2 \eta^3)$ time complexity. Note that these estimates are for naive matrix multiplication. Regardless, the cubic scaling in $\eta$ dominates the time-complexity for observables with many QoIs. For example, to build a GP surrogate model for the density of a noble gas ($\eta = 1$) with Lennard-Jones interactions (dim($\boldsymbol{\theta}$) = 2) would give a training set matrix of ($2N \times 3)$. Similarly, a surrogate model for an infrared spectrum of water from 600-4000 cm$^{-1}$ at a resolution of $4$ cm$^{-1}$ ($\eta = 850$) estimated with a 3 point water model of Lennard-Jones type interactions (dim($\boldsymbol{\theta}$) = 6) would generate a training set matrix of size (850$N$ $\times$ 7). Clearly, the complexity of the output QoI causes a significant increase in the computational cost of the matrix operations.

\subsection{The Local Gaussian Process Surrogate Model}

The time-complexity of the training-kernel matrix inversion and the matrix product can be substantially reduced by fragmenting the full Gaussian process of eq \eqref{eq:surrogate} into $\eta$ Gaussian processes. This method is also referred to as the subset of regressors approximation \cite{silverman_aspects_1985} and is considered a "greedy" approximation \cite{rasmussen_gaussian_2006}. Under this construction, an individual $GP_k$ is trained to map a set of model parameters to an individual QoI,

\begin{equation}
    \mathbb{E}[GP_k] : \boldsymbol{\theta} \mapsto S(r_k)
\end{equation}

\noindent where $\mathbf{r}$ is no longer an input parameter. The training set matrix, $\mathbf{\hat{X'}}$, is now a ($N$ $\times$ dim($\boldsymbol{\theta}$)) matrix, 

\begin{equation}\label{eq:subsurrogate_training}
    \mathbf{\hat{X'}} = 
        \begin{bmatrix}
        \theta_{1,1} & \theta_{2,1} & \hdots\\
        \theta_{1,2} & \theta_{2,2} & \hdots\\
        \vdots & \vdots & \vdots\\
        \theta_{1,N} & \theta_{2,N} & \hdots\\
        \end{bmatrix}
\end{equation}

\noindent while the training set observations, $\mathbf{\hat{Y'}}_k$, is a ($N$ $\times$ 1) column vector of the QoIs from the training set at $r_k$,

\begin{equation}\label{eq:subsurrogate_observation}
    \mathbf{\hat{Y}'}_k = [S(\boldsymbol{\theta}_1,r_k), ..., S(\boldsymbol{\theta}_N,r_k)]^T
\end{equation}

\noindent where $S(\boldsymbol{\theta}_j,r_k) = y(\boldsymbol{\theta}_j,r_k) - \mathbf{\mu}_{LGP, k}^{prior}(r_k)$ and $k$ indexes over IVs. The LGP prior mean $\mathbf{\mu}_{LGP, k}^{prior}(r_k)$ is now,

\begin{equation}\label{eq:subsurrogate_mean}
    \mathbf{\mu}_{LGP, k}^{prior} := [\mu(\boldsymbol{\theta}_1,r_k), ..., \mu(\boldsymbol{\theta}_N,r_k)]^T
\end{equation}

\noindent such that $\mu(\boldsymbol{\theta}_j,r_k)$ is the GP prior mean for parameter $\boldsymbol{\theta}_j$ at $r_k$. The squared-exponential kernel function is now,

\begin{equation}\label{eq:LGPkernel}
    K_{mn} = \alpha^2 \exp\bigg(- \sum_{o=1}^{\text{dim}(\boldsymbol{\theta})}\frac{(\theta_{o,m} - \theta_{o,n})^2}{2\ell_{\theta_o}^2} \bigg).
\end{equation}

\noindent The LGP surrogate model expectation for the observable at $r_k$, at a new set of parameters, $\boldsymbol{\theta}^*$, is just the expectation of the $k^{th}$ Gaussian process given the training set data,

\begin{equation}\label{eq:subsurrogate}
    S_{loc}^*(r_k|\boldsymbol{\theta}^*) = \mathbb{E}[\textit{GP}_k(\boldsymbol{\theta}^*)] = \mathbf{K}_{\boldsymbol{\theta}^*,\mathbf{\hat{X'}}} [\mathbf{K}_{\mathbf{\hat{X'}}, \mathbf{\hat{X'}}} + \sigma_{noise}^2 \mathbf{I}]^{-1} \mathbf{\hat{Y'}}_k
\end{equation}

\noindent We then just combine the local results from the subset of $\eta$ GPs to obtain a prediction for the difference between the model and LGP prior mean,

\begin{equation}\label{eq:subQOI}
    S_{loc}^*(\mathbf{r}|\boldsymbol{\theta}^*) = [S_{loc}^*(r_1|\boldsymbol{\theta}^*), ..., S_{loc}^*(r_\eta|\boldsymbol{\theta}^*)]^T.
\end{equation}

\noindent and subsequently add back the LGP prior mean to obtain the estimated QoI, $y_{loc}^*(\mathbf{r}|\boldsymbol{\theta}^*) = S_{loc}^*(\mathbf{r}|\boldsymbol{\theta}^*) + \mathbf{\mu}_{LGP, k}^{prior}(\boldsymbol{\theta}^*,\mathbf{r})$.

By reducing the dimensionality of the relevant matrices, the time complexity of the matrix calculations are drastically reduced compared to a standard GP. The single step inversion of the training-training kernel matrix is now of $\mathcal{O}(N^3)$ time complexity while the $\eta$ step (1 $\times$ $N$) $\times$ ($N$ $\times$ $N$) $\times$ ($N$ $\times$ 1) matrix products are reduced to $\mathcal{O}(N^2 \eta)$ time complexity. If the number of training samples, $N$, the number of IVs, $\eta$, and the number of model evaluations, $G$, are equal between the full and LGP algorithms, then a LGP approximation reduces the evaluation time complexity in a standard GP from cubic-scaling, $\eta^3$, to embarrassingly parallelizable linear-scaling, $\eta$.

In summary, a local Gaussian process is an approximation in which the QoIs are modeled as independent random variables, each described by their own Gaussian process. This amounts to assuming that the random variables are stochastically independent. For time-independent data including scattering measurements and spectroscopy, this approximation is appropriate since each observation is an independent measurement at each independent variable. Finally, it is well-established that low rank approximations of Gaussian processes can compromise the accuracy of the estimated uncertainty, so the use of LGP regressors should be carefully scrutinized based on the risk/consequences of misrepresenting the resulting functional distributions.

Complex experimental observables can be reconstructed by this set of LGPs through a series of relatively straightforward matrix operations with linear time-complexity in the number of IVs. Furthermore, the LGP has all of the primary advantages of Bayesian methods, including built-in UQ and analytical derivatives and Fourier transforms. In the following section, we demonstrate the computational enhancement and accuracy of the LGP approach by modeling the RDF of neon at 42$K$. The LGP surrogate model is then implemented within a Bayesian framework to exemplify the power of UQ/P for molecular modeling.  

\section{A Local Gaussian Process Surrogate for the RDF of Liquid Ne}

To explore the computational advantages of LGP surrogate models for Bayesian inference, we studied the experimental RDF of liquid Ne \cite{bellissent-funel_neutron_1992} under a ($\lambda$-6) Mie fluid model. The ($\lambda$-6) Mie force field is a flexible Lennard-Jones type potential with variable repulsive exponent, 

\begin{equation}
    v^{Mie}_2(r) = \frac{\lambda}{\lambda-6}\bigg(\frac{\lambda}{6}\bigg)^{\frac{6}{\lambda-6}} \epsilon \bigg[ \bigg(\frac{\sigma}{r}\bigg)^\lambda - \bigg(\frac{\sigma}{r}\bigg)^6 \bigg]
\end{equation}

\noindent where $\lambda$ is the short-range repulsion exponent, $\sigma$ is the collision diameter ($\si{\angstrom}$), and $\epsilon$ is the dispersion energy (kcal/mol) \cite{mie_zur_1903}.  

MD simulations were performed from a Sobol sampled set spanning a prior range based on existing force field models \cite{vrabec_set_2001, mick_optimized_2015, shanks_transferable_2022} ($\lambda = [6.1,18]$, $\sigma = [0.88, 3.32]$, and $\epsilon = [0, 0.136]$) to generate a RDF training set matrix of the form in eq \ref{eq:subsurrogate_training}. Prior parameter ranges were selected so that training samples were restricted to the liquid regime of the ($\lambda$-6) Mie phase diagram \cite{widom_new_1970, ramrattan_corresponding-states_2015}. A sequential sampling approach was used in which we Sobol sample the prior range of parameters, calculate the training sample with the best-fit to the experimental data (lowest root mean squared error), center the new space on this training sample, and then narrow the sample range around this center point by a user selected ratio $\gamma$. This procedure was repeated three times with 320 samples per round (960 total training simulations) with $\gamma = 0.8$. This ratio was selected so that the final range would span >3 standard deviations of the posterior distributions estimated in prior literature \cite{angelikopoulos_bayesian_2012, mick_optimized_2015}. Subsequently, 320 test simulations were randomly sampled from the final range and used to determine whether or not the surrogate model provides accurate model predictions. A visualization of this procedure is provided in the Supporting Information. 

The number of observed points $\eta$ in the radial distribution function was calculated by dividing the reported $r_{max} - r_{min} \approx 15.3$ by the effective $r$-space resolution given by, $\Delta r = \pi/Q_{max}$, where $\Delta r = 0.21$ $\si{\angstrom}$ for reported $Q_{max} = 15$ $\si{\angstrom}^{-1}$. This relation indicates that the appropriate number of observed independent $r$-values in the RDF is $\eta = 73$.

The training set matrix and training observation matrix were then constructed from the 960 training samples according to eqs \eqref{eq:subsurrogate_training} and \eqref{eq:subsurrogate_observation}, respectively. As a prior mean, we selected the RDF determined analytically from the dilute limit potential of mean force (PMF),

\begin{equation}
    \mathbf{\mu}_{PMF, k}^{prior}(\boldsymbol{\theta}_j,r_k) := g(\boldsymbol{\theta}_j,r_k) = \exp{[-\beta V(\boldsymbol{\theta}_j,r_k)]}
\end{equation}

\noindent where $g(\boldsymbol{\theta}_j,r_k)$ and $V(\boldsymbol{\theta}_j,r_k)$ are the analytical dilute limit RDF and ($\lambda$-6) Mie potential for parameters $\boldsymbol{\theta}_j$ at $r_k$, respectively. A PMF prior mean yields physically realistic short-range ($g(r) = 0$) and long-range behavior ($g(r) \to 1$). The PMF prior had improved RMSE compared to an ideal gas prior ($\forall r \in \mathbb{R}_0^+$, $g(r) = 1$), but this difference did not significantly impact the Bayesian posterior estimate (see Supporting Information). Finally, LGP hyperparameter optimization was performed using brute force to maximize the log-marginal likelihood \cite{sundararajan_predictive_2001} over the training set.

Quantitative analysis of model sensitivity can be performed with probabilistic derivatives of the QoI with respect to model parameters (see Supporting Information) and subsequently related to temperature derivatives of radial distribution functions \cite{piskulich_temperature_2020}.

\subsection{Computational Efficiency and Accuracy}

Now that we have constructed the training set matrix, we simply evaluate the expectation at each $r_k$ according to eq \eqref{eq:subsurrogate} and combine the results into a single array as in eq \eqref{eq:subQOI}. The average computational time to invert the training set matrix and evaluate the surrogate model for both a standard GP and LGP are shown below in Table \ref{tab:speed}. The LGP surrogate accelerates the RDF evaluation time compared to molecular dynamics by a factor of 1,700,000 for the $\eta = 73$ independent variable QoI with 960 training simulations. This 6 orders-of-magnitude speed-up beats a standard GP by 3 orders-of-magnitude (2141x). With respect to the training-training kernel matrix inversion, the LGP wins out on the standard GP by a factor of 31,565.

\begin{table}
\centering
\caption{Average relative time and speed-up to QoI evaluation and training set matrix inversion for a standard and local Gaussian process for 960 training samples and a RDF with $\eta = 73$ points.}
\begin{tabular}{| l | c | c | c | c|}
\hline
\textrm{Model}&
\textrm{QoI Eval. Time (s)}&
\textrm{Speed Up ($t/t_{sim}$)}&
\textrm{Inv. Time (s)}\\
\hline
Simulation  & 1,251  & 1      & - \\
GP          & 1.52   & 822     & 355\\
LGP         & 0.0007 & 1,760,267 & 0.01\\
\hline
\end{tabular}
\label{tab:speed}
\end{table}

In summary, the LGP significantly accelerates both computational bottlenecks for Gaussian process surrogate modeling; namely, the training set matrix inversion and surrogate model evaluation time. Of course, the exact speed-ups depend on numerous factors including the number of IVs $\eta$, the number of training samples used to construct the training set matrix $N$, the level of code parallelization, and hyperparameter optimization procedure. Which step is rate limiting depends on the surrogate modeling application. For instance, if the surrogate model doesn't need to be evaluated a large number of times, the training set generation, matrix inversion and hyperparameter optimization will be the rate limiting steps. On the other hand, applications that require a large number of model evaluations, such as uncertainty quantification and propagation, result in the surrogate model evaluation time being rate limiting. Typically, designing a surrogate model is only necessary in the latter case.  

Clearly the LGP is fast, but is it accurate? In other words, does the LGP provide QoI predictions that are within a reasonable level of accuracy to serve as a true surrogate model for the molecular dynamics predictions? To evaluate the accuracy of the local predictions, a test set of 320 ($\lambda$-6) Mie parameters was randomly sampled from the final range of the sequential sampling method (see Supporting Information) and the RMSE computed between simulated and LGP predicted radial distribution functions along all radial positions, $r$. The results are summarized below in Figure \ref{fig:rmse}.

\begin{figure}[H]
    \centering
    \includegraphics[width = 16 cm]{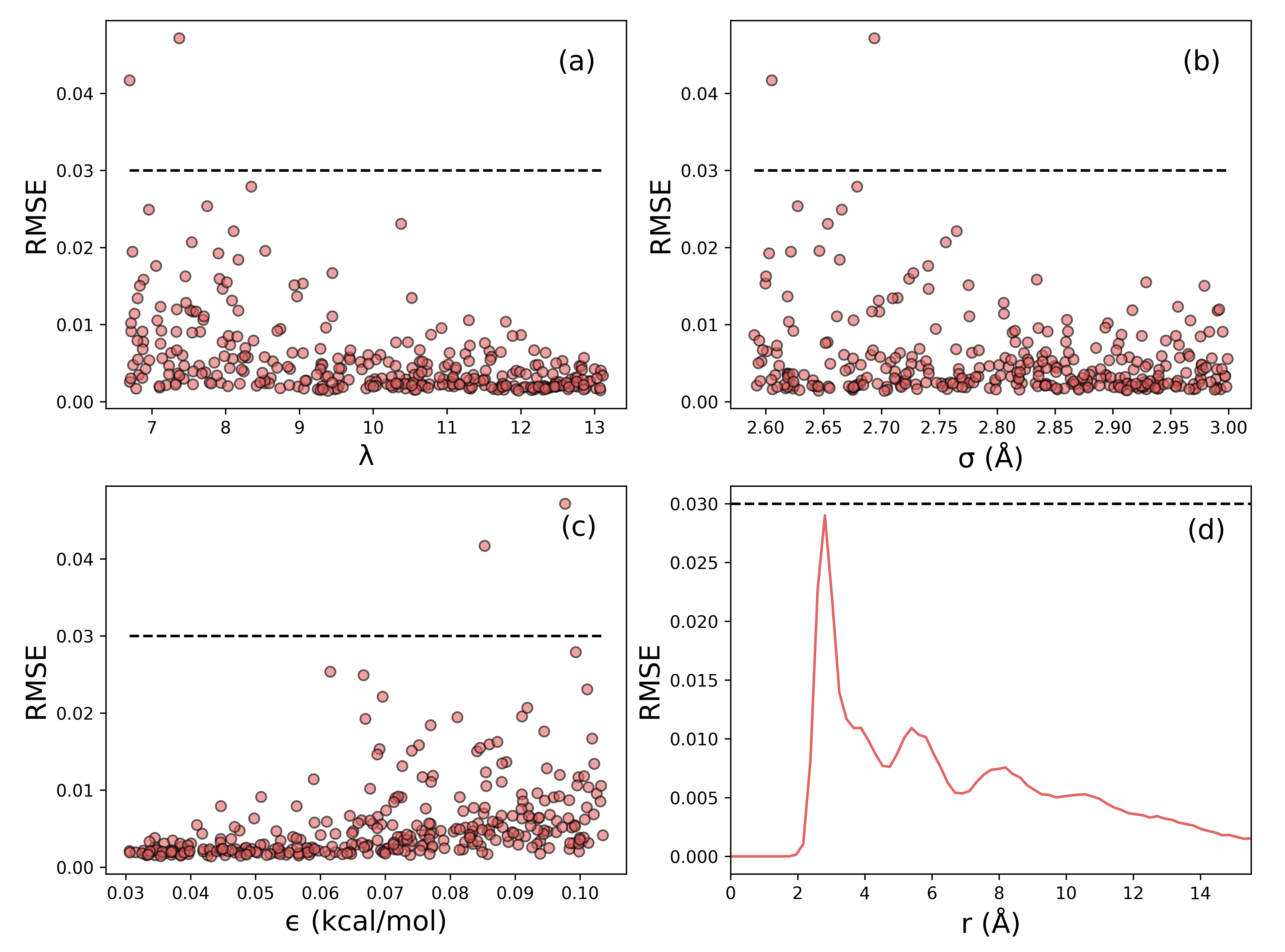}
    \caption{(a)-(c) Test set samples over each parameter plotted against the RMSE between simulated and LGP data. (d) Average RMSE over the 320 test set samples as a function of $r$. The dashed line represents the reported error from the experiment \cite{bellissent-funel_neutron_1992}.}
    \label{fig:rmse}
\end{figure}

\noindent The RMSE for all radial positions is less than 0.03, which is excellent considering that this error is smaller than the reported experimental uncertainty ($\sim$0.03). Of course, the acceptable RMSE over the QoI is user-defined and largely subjective based on the surrogate model application, but can be improved with additional training and hyperparameter optimization if necessary (an example is included in the Supporting Information). 

\subsection{Learning from the Ne RDF Surrogate Model with Bayesian Analysis}

Our fast and accurate LGP surrogate model now allows us to explore the underlying probability distributions on the ($\lambda$-6) Mie parameter space. This example is provided to show how one can use Bayesian analysis to learn about correlations and relationships between model parameters as well as model adequacy. This analysis can provide robust insight into the nature of the model and provide quantifiable evidence for whether or not the model is appropriate for a target application.    
Bayesian inference yields a probability distribution function over the model parameters called the joint posterior probability distribution. The maximum of the joint posterior, referred to as the \textit{maximum a posteriori} (\textit{MAP}), represents the set of parameters with the highest probability of explaining the given experimental data. In force field design, the \textit{MAP} would be an appropriate choice for an optimal set of model parameters. However, the power of the Bayesian approach lies in the fact that, not only can we identify the optimal parameters, but we can also examine the probability distribution of the parameters around these optima. For instance, the width of the distribution provides evidence for how important a parameter in the model is for representing the target data. For a given parameter, a wide distribution indicates that the parameter has little influence on the model prediction. On the other hand, a narrow distribution indicates that the parameter is critical to the model prediction. Additionally, the joint posterior may exhibit multiple peaks, or modes. A multimodal joint posterior suggests that there are multiple sets of model parameters that reproduce the target data, which may be a symptom of model inadequacy. Finally, the symmetry of the distribution provides information on relationships and correlations between parameters, providing a framework to diagnose subtle relationships that may otherwise go unnoticed.

\begin{figure}[H]
    \centering
    \includegraphics[width = 16cm]{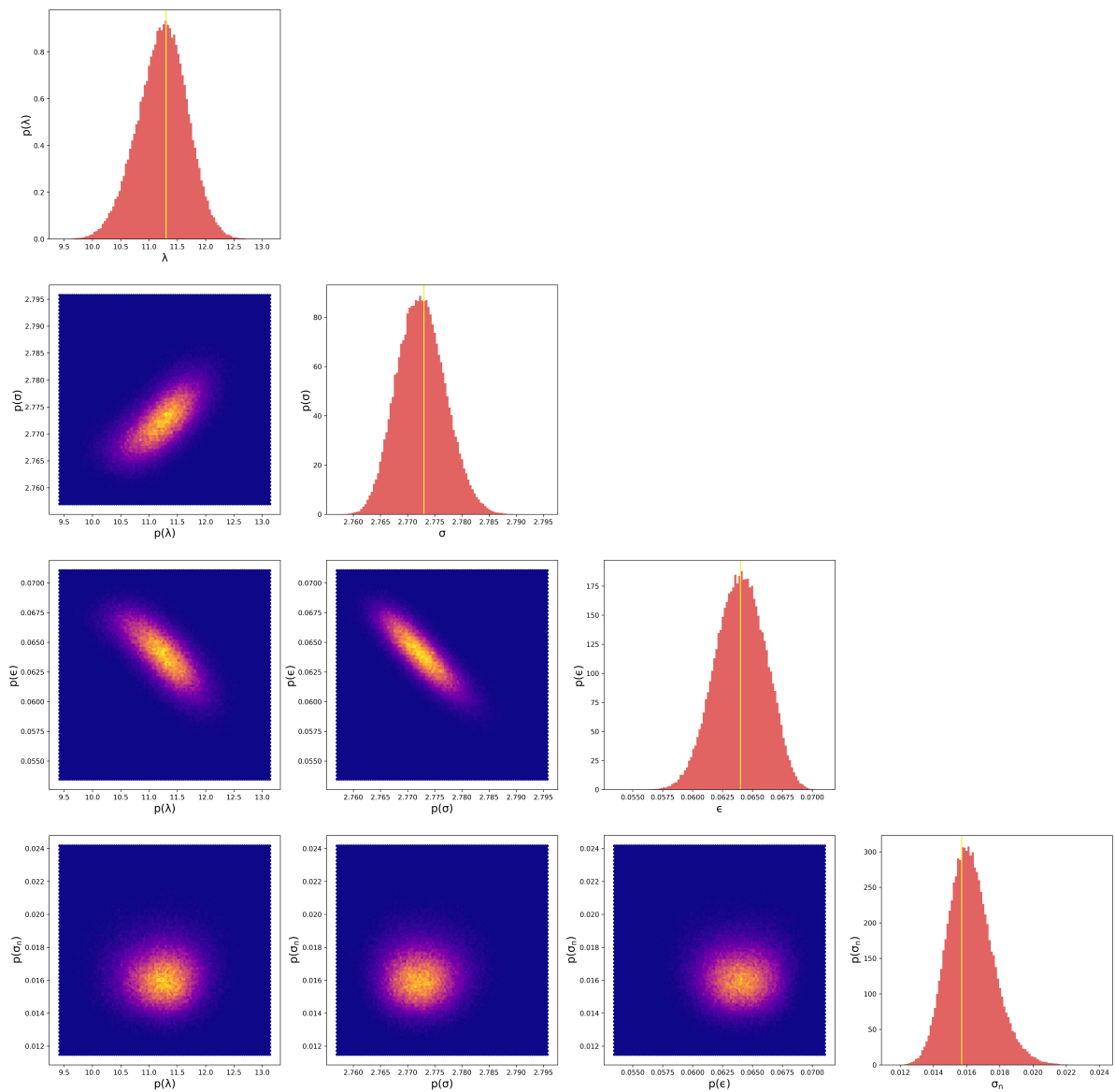}
    \caption{(Diagonal) 1D marginal distributions for the ($\lambda$-6) Mie fluid parameters. Prior distributions are not depicted since they are flat lines near 0 probability. Yellow vertical lines represent the \textit{maximum a posterior} (\textit{MAP}) estimate. (Off-Diagonal) 2D marginal histograms showing parameter correlations.}
    \label{fig:posterior}
\end{figure}

Usually, the joint posterior distribution is a high-dimensional quantity that cannot be visualized directly. However, we can visualize the joint posterior along one dimension by integrating out the contributions over all other parameters. The resulting distributions are called marginal distributions. Marginal distributions computed over the ($\lambda$-6) Mie potential parameters optimized to the RDF of liquid Ne are shown in Figure \ref{fig:posterior}.

\noindent For each parameter, the resulting marginal posterior distributions are unimodal and symmetric. This result is not surprising in the context of recent results that show iterative Boltzmann inversion, which is a maximum likelihood approach to the structural inverse problem, is convex for Lennard-Jones type fluids \cite{hanke_well-posedness_2018}. Observing the 2D marginal distributions in Figure \ref{fig:posterior}, we can also see that each of the parameters are correlated to one other. For example, the negative correlation between $\sigma$ and $\epsilon$ suggests that increasing the size of the particle should be accompanied by a decrease in the effective particle attraction. Conceptually this makes sense, if the particles are larger, then they would need to have a weaker attractive force to give the same atomic structure. This result is consistent with Bayesian analysis on liquid Ar \cite{angelikopoulos_bayesian_2012}. The nuisance parameter distribution shows that the unknown standard deviation between the LGP surrogate model and the experimental data is around 0.016.

One surprising characteristic of the posterior distribution is that it is extremely narrow. Recall that narrow distributions indicate that the parameters are important, or have tight control, over the model quality-of-fit to the experimental data. From our Bayesian analysis, we can therefore confidently conclude that detailed interatomic force information is contained within the experimental RDF. This observation is in stark contrast to over 60 years of prior literature which has unanimously asserted that only the excluded volume or collision diameter can be ascertained from experimental scattering data \cite{clayton_neutron_1961, jovari_neutron_1999, hansen_theory_2013}. In fact, the Bayesian analysis shows that it is possible to determine values for $\lambda$, $\sigma$, and $\epsilon$ within $\pm 2$, $\pm 0.02 \si{\angstrom}$, and $\pm 0.0075$ kcal/mol with 95\% certainty. This result leads to two important conclusions: (1) Scattering data can effectively constrain the force field model parameter space and (2) the data must be sufficiently accurate to do so. These results provide evidence that scattering data could be invaluable to inform accurate force fields, particularly for structure and self-assembly applications.

The joint posterior can also be used for model parameter selection given the experimental observation. Specifically, the optimal parameters are given by the \textit{MAP}, corresponding to the maximum of the joint posterior distribution. The \textit{MAP} is presented in Table \ref{tab:params} along with two other existing force fields for liquid Ne.

\begin{table}[H]
\centering
\caption{Summary of ($\lambda-6)$ Mie potential parameters optimized for Ne. Values for the repulsive exponent parameter are rounded to the nearest integer.}
\begin{tabular}{| l | c | c | c | c | c |}
\hline
\textrm{Force Field}&
\textrm{QoI}&
\textrm{$\lambda$}&
\textrm{$\sigma$ ($\si{\angstrom}$)}&
\textrm{$\epsilon$ (kcal/mol)}\\
\hline
Mick (2015) & VLE & 11 & 2.794 & 0.064\\
SOPR (2022) & RDF & 11 & 2.778 & 0.063\\
This Work   & RDF & 11 & 2.773 & 0.064\\
\hline
\end{tabular}
\label{tab:params}
\end{table}

\noindent The estimated Mie parameters are in agreement with the Mick \cite{mick_optimized_2015} and structure optimized potential refinement (SOPR) \cite{shanks_transferable_2022} models. This result confirms that the radial distribution function contains sufficient information to determine transferable force field parameters in simple liquids. 

Some interesting questions arise considering that both the Mie fluid model and SOPR, which is a probabilistic iterative Boltzmann method for experimental scattering data, give similar predictions for the structure-optimized potentials. The key difference between the Bayesian optimization performed in this work and SOPR is that the former is parametric while the latter is non-parametric, both of which have strengths and weaknesses. Specifically, parametric models are less complex but may not be flexible enough to describe subtle details of the experimental observation. On the other hand, non-parametric models can describe nuanced experiments but may over-fit to non-physical features of the data. It is then natural to wonder: Is a ($\lambda$-6) Mie model adequate to describe the experimental scattering data? Or does the scattering data complexity necessitate the use of non-parametric iterative potential refinement techniques like SOPR?

\begin{figure}[H]
    \centering
    \includegraphics[width = 11cm]{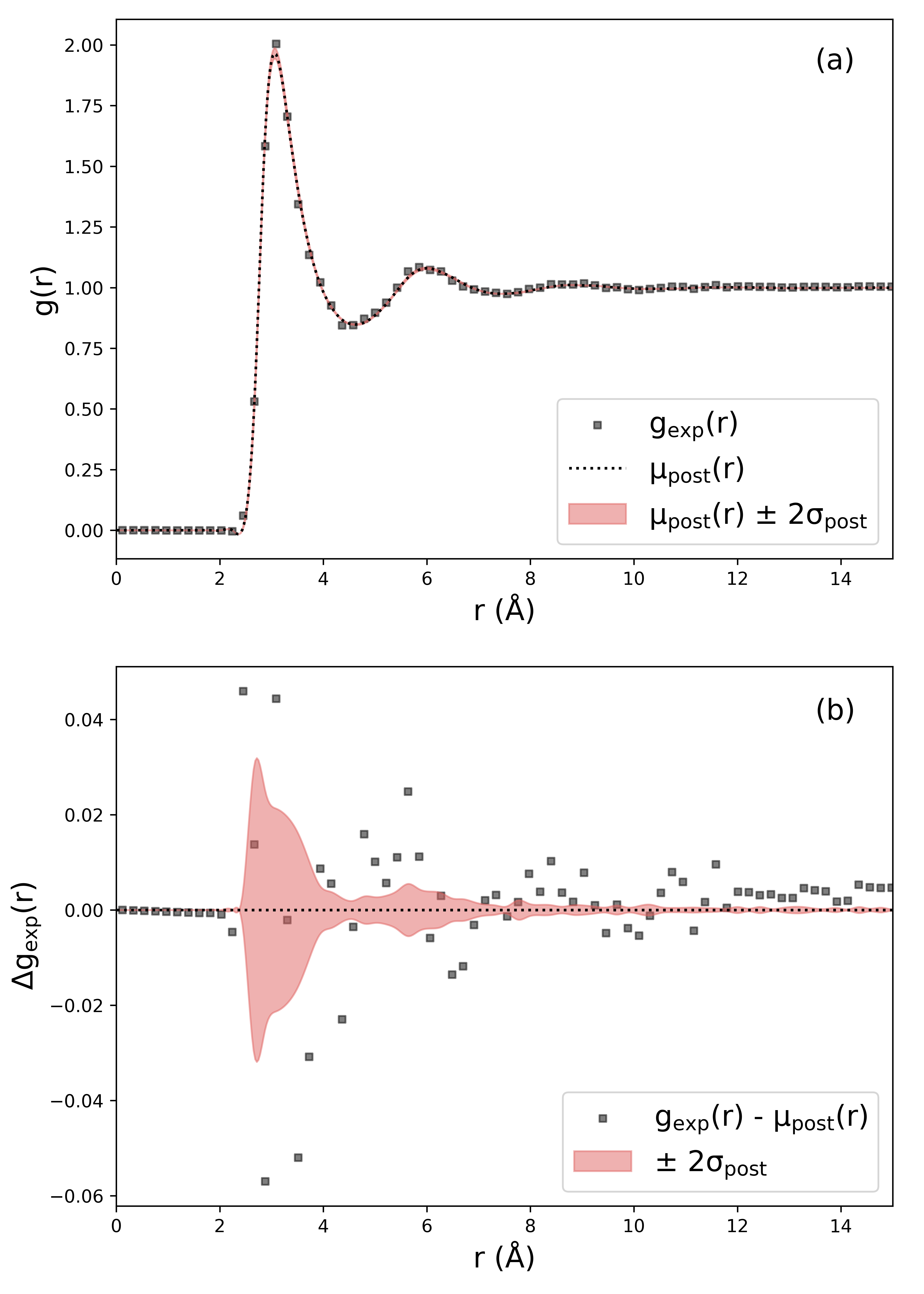}
    \caption{(a) RDF mean and credibility interval propagated from the parameter uncertainty quantified with Bayesian inference. (b) Residual analysis comparing the experimental data with the posterior predictive distribution.}
    \label{fig:post_pred}
\end{figure}

We can investigate the first question of model adequacy by propagating parameter uncertainty through the LGP to construct a distribution of RDF predictions - referred to as the posterior predictive. The posterior predictive can be estimated by evaluating the LGP for all MCMC samples and computing the mean,

\begin{equation}
        \mathbb{E}[{S}_{loc}^*(\mathbf{r}_k)] \approx \frac{1}{N}\sum_{i=1}^N {S}_{loc}^*(\mathbf{r}_k|\boldsymbol{\theta}_i)
\end{equation}

\noindent and variance,

\begin{align}
    \mathbb{V}[{S}_{loc}^*(\mathbf{r}_k)] \approx \frac{1}{N}\sum_{i=1}^N ({S}_{loc}^*(\mathbf{r}_k|\boldsymbol{\theta}_i) - \mathbb{E}[{S}_{loc}^*(\mathbf{r}_k)])^2 
\end{align} 

\noindent of the resulting QoI predictions. Recall that the nuisance parameter distribution is also sampled to account for unknown uncertainties in the LGP surrogate model and experimental data. The posterior predictive therefore quantifies of how accurately we know the QoI given experimental, model, and parametric uncertainty estimated with Bayesian inference. If the model is adequate, the Bayesian credibility interval ($\mu \pm 2 \sigma$) should contain approximately 95\% of the experimental data. The posterior predictive and residuals ($g_{exp}(r) - \mu(r)$) estimated for the liquid Ne RDF are shown in Figure \ref{fig:post_pred}.

\noindent Clearly, the agreement between the posterior predictive mean and the experimental data is excellent. However, the residuals often lie outside of the $2\sigma_{post}$ credibility interval. These differences between the experiment and model could be explained by a number of different factors, including errors arising from Fourier transform truncation, background scattering corrections or model inadequacy, among others. However, without rigorous uncertainty quantification on the experimental scattering data, it is currently not possible to determine which factor or combination of factors results in the model disagreement. We argue that this knowledge gap necessitates rigorous UQ/P studies on scattering data as well as iterative potential refinement methods. Combining these approaches with Bayesian inference on molecular dynamics models could then shed light on what physical interactions can be learned from scattering experiments.

In summary, we have shown that a LGP surrogate model enables rapid and accurate uncertainty quantification and propagation with Bayesian inference. We then showed how the posterior distribution is an indispensable tool to learn subtle relationships between model parameters, identify how important each model parameter is to describe the outcome of experiments, and quantify our degree of belief that our model adequately describes our observations. The power of Bayesian inference is evident.

\section{Conclusions}

We have shown that local Gaussian process surrogate models trained on an experimental RDF of liquid neon improves the computational speed of QoI prediction 1,760,000-fold with exceptional accuracy from only 960 training simulations. The 3 orders-of-magnitude evaluation time speed-up for a local versus standard Gaussian process was shown to accelerate Bayesian inference without the need for advanced sampling techniques such as on-the-fly learning. Furthermore, since the LGP linearly scales with the number of output QoIs, significantly higher speed-ups are expected for more complex data, such as infrared spectra or high resolution scattering experiments, or for multiple data sources simultaneously (\textit{e.g.} scattering, spectra, density, diffusivity, etc). We conclude that local Gaussian processes are an accurate and reliable surrogate modeling approach that can accelerate Bayesian analysis of molecular models over a broad array of complex experimental data. 

\section{Acknowledgements}

We would like to thank Dr. J\"urgen D\"olz for helpful discussions on the implementation of local Gaussian processes, hyperparameter optimization and uncertainty quantification.  Dr. Valeria Molinero for providing feedback on writing and scientific communication in preparation of the manuscript. Finally, we thank Sean T. Smith and Philip J. Smith for their guidance on Bayesian methodology. This study is supported by the National Science Foundation Award No. CBET-1847340. The support and resources from the Center for High Performance Computing at the University of Utah are gratefully acknowledged.  

\section{Author Contributions}

BL Shanks - conceptualization, formal analysis, code development, manuscript writing and preparation. HW Sullivan - conceptualization, formal analysis, code development, manuscript preparation. AR Shazed - molecular simulations. MP Hoepfner - conceptualization, funding acquisition, manuscript preparation.

\printbibliography

\section{Appendix}
\appendix
\input{si.tex}

\typeout{get arXiv to do 4 passes: Label(s) may have changed. Rerun}

\end{document}

%% file: si.tex
\section{Molecular Dynamics Simulation of Mie Fluids}

Computer generated radial distribution functions (RDFs) were calculated using molecular dynamics (MD) simulations in the HOOMD-Blue package \cite{anderson_hoomd-blue_2020}. Simulations were initiated with a lattice configuration of 864 particles and compressed to a reduced density of $\rho = 0.02477$ atom/$\si{\angstrom}^3$ and thermal energy $T = 42.2$ K. The HOOMD NVT integrator was used for a 0.25 nanosecond equilibration step and a 0.25 nanosecond production step (dt = 0.5 femtosecond). Potentials were truncated at $3\sigma$ with an analytical tail correction, and RDFs were calculated using the Freud package \cite{ramasubramani_freud_2020}.

\section{Training and Test Set Generation}

The first step to design a LGP surrogate model is to generate a training set of model input parameter input and QoI outputs. To generate the training set, we need dense samples of model parameters in the region of the parameter space that well-represents the target experimental data. In general, it is not known \textit{a priori} where this region is, particularly if there is no prior knowledge of what model parameters are best with respect to an experimental observation, $\mathbf{y}$. However, there are parameter regions that we can exclude \textit{a priori} based on the physics of the ($\lambda$-6) Mie fluid. For instance, Ne is a liquid at the experimental thermodynamic conditions, so we can use well-established ($\lambda$-6) Mie fluid phase diagrams and vapor-liquid transitions \cite{widom_new_1970} to restrict the parameter ranges to the liquid phase only. Specifically, given a fixed temperature (T = 42.2K) and density ($\rho$ = 0.024 $\si{\angstrom}^{-3}$), it is trivial to determine the $\sigma$ and $\epsilon$ parameter ranges reported in the manuscript via relations for the scaled temperature ($T^* = k_bT/\epsilon$) and scaled density ($\rho^* = \rho \sigma^3$). The parameter ranges determined using the Mie fluid phase diagram are presented in Table \ref{tab:ranges}. Restricting the parameter to physically justified ranges is important to avoid a "garbage in, garbage out" scenario for an LGP surrogate model. Given this prior range, we then performed the sequential sampling approach outlined in the manuscript. A visualization of this procedure is shown below in Figure \ref{fig:trainingset}.

\begin{table}
\centering
\caption{Estimated boundaries for physics-constrained prior space based on the ($\lambda$ - 6) Mie fluid phase diagram. $m = 6$ is the attractive tail exponent of the ($\lambda$ - 6) Mie potential. $*$) The maximum $\lambda$ was selected to be substantially larger than previously reported values \cite{vrabec_set_2001, mick_optimized_2015, shanks_transferable_2022}.}
\begin{tabular}{| c | l | l | l | l |}
\hline
\textrm{Param.}&
\textrm{Min.}&
\textrm{Min. Criteria}&
\textrm{Max.}&
\textrm{Max. Criteria}
\\
\hline
$\lambda$  & 6.1  & $m=6 \implies \lambda>6$  &  18 & Literature$^*$ \\
$\sigma$   & 2.55 & Vapor-Liquid Equil. &  3.32 & Solid-Liquid Equil.  \\
$\epsilon$ & 0.00 & $\epsilon<0$ undefined  &  0.136 & Vapor-Solid Equil. \\
\hline
\end{tabular}
\label{tab:ranges}
\end{table}

\begin{figure}
    \centering
    \includegraphics[width = 16cm]{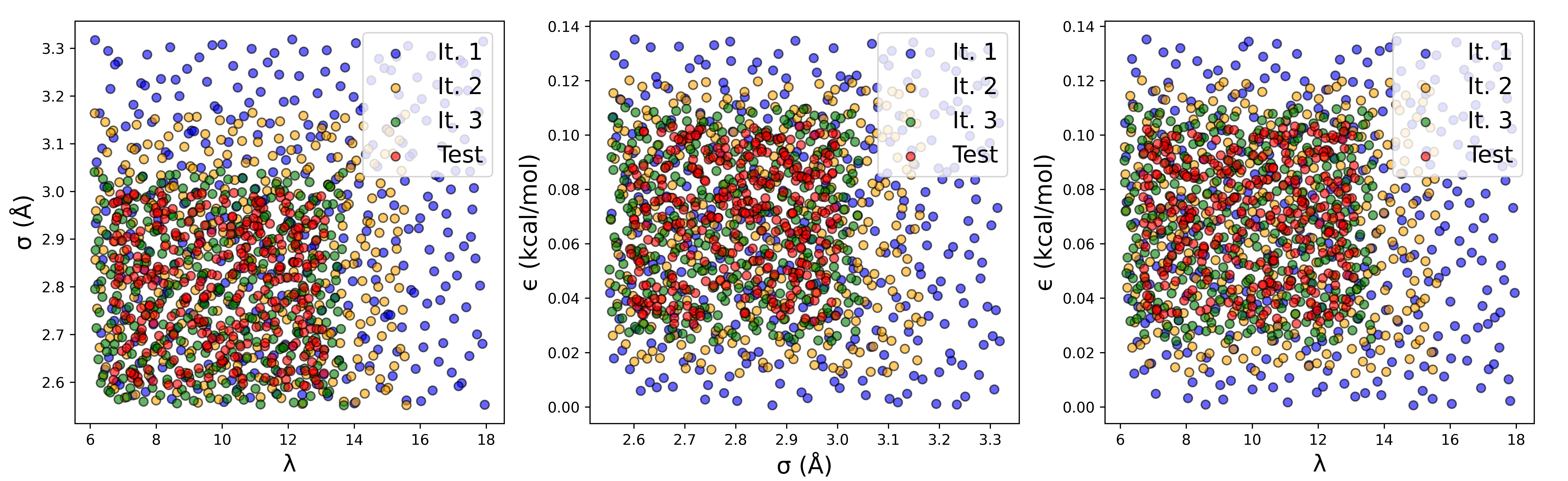}
    \caption{2D training and sample parameter set used to train and test the LGP surrogate model.}\label{fig:trainingset}
\end{figure}

\subsection{The Gaussian Process Prior Mean}

\begin{figure}
    \centering
    \includegraphics[width = 16cm]{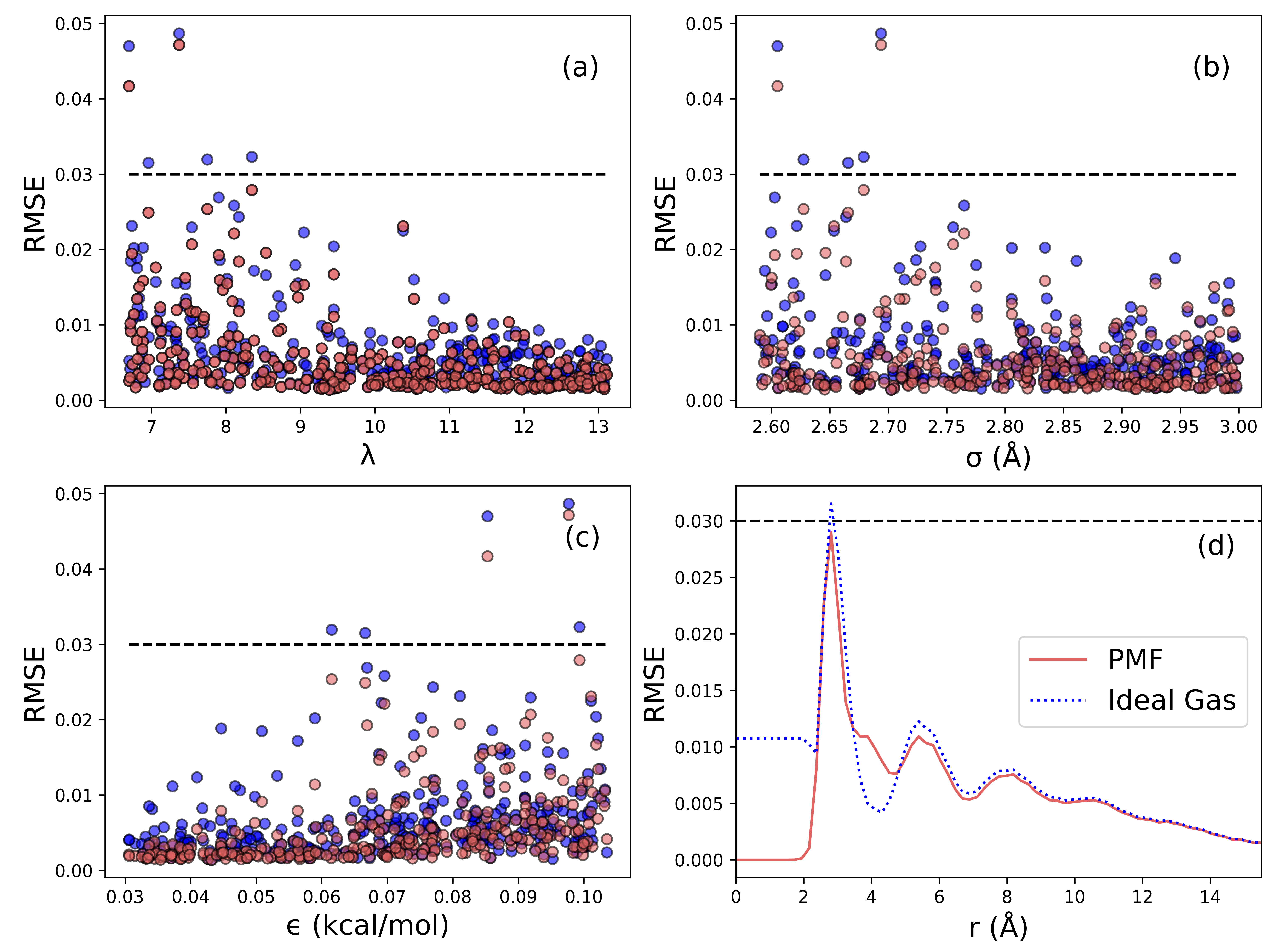}
    \caption{RMSE as a function of $r$ for an ideal gas (blue) and potential of mean force (red) prior.}
    \label{fig:rmse2}
\end{figure}

In this manuscript, an analytical solution for the RDF based on the dilute limit potential of mean force (PMF) was used as the GP prior mean. This choice is appropriate as an RDF prior since it will have the same features that we expect a liquid RDF to have, \textit{i.e.} RDF values of zero at low $r$ and a long-range tail that asymptotically approaches unity. However, note that even a prior guess that doesn't encode this information can still produce accurate LGP surrogate models for RDFs. For example, in Figure \ref{fig:rmse2} we can see that an ideal gas RDF prior, which amounts to approximating that the RDF is unity everywhere ($g^{IG}(r) = 1$), can still be learned by the local Gaussian processes with RMSE values close to the more physically justified PMF.

\begin{figure}
    \centering
    \includegraphics[width = 16cm]{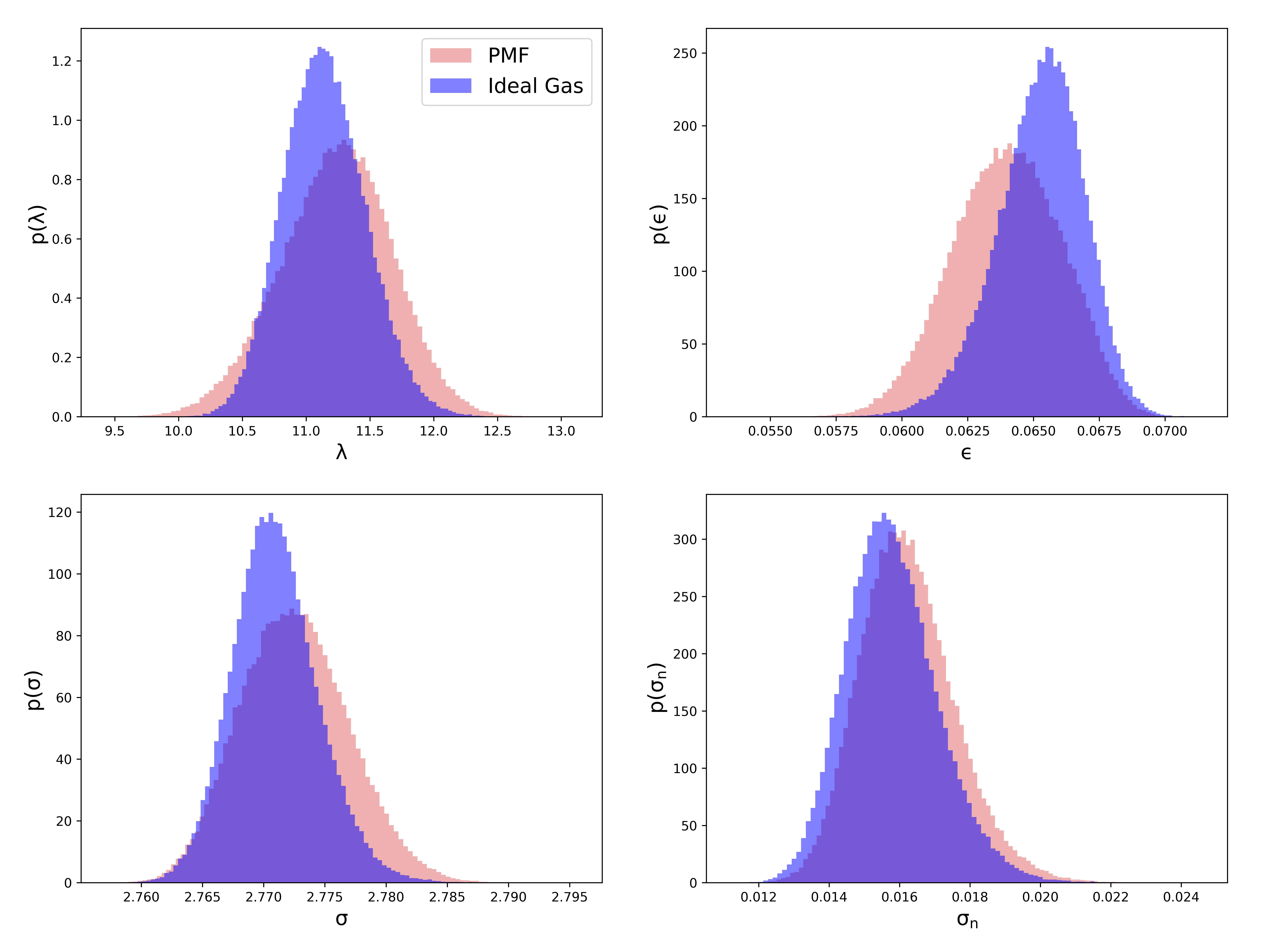}
    \caption{Marginal posteriors for the ideal gas PMF (red) and ideal gas (blue) priors.}
    \label{fig:post_compare}
\end{figure}

\begin{figure}
    \centering
    \includegraphics[width = 12cm]{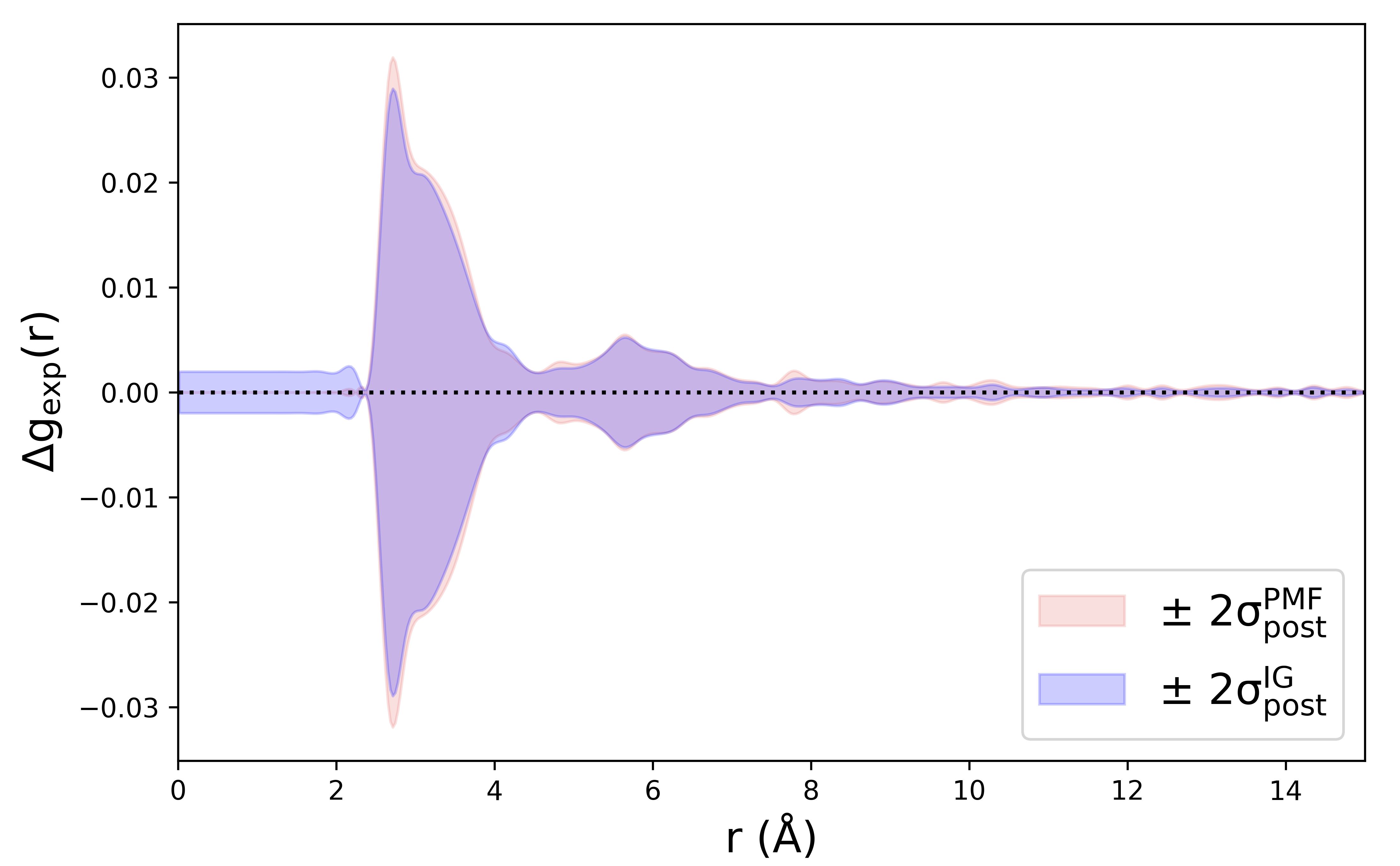}
    \caption{Posterior predictives for the ideal gas PMF (red) priors ideal gas (blue) priors.}
    \label{fig:postpredictive_compare}
\end{figure}

Clearly, the RMSE along $r$ is a similar magnitude for the ideal gas and PMF prior, but the $r$-dependent behavior is noticeably different. For the ideal gas prior, we see that there is high RMSE at low $r$, which is inconsistent with our intuition for a liquid RDF due to the excluded volume of atoms. What is occurring here is that the GP estimate is being "pulled" towards the prior at low $r$. On the other hand, the PMF prior exhibits behavior in line with our physical intuition; namely, near zero error at $r$ values smaller than the relative diameter of the atom. Perhaps surprisingly, we see in Figure \ref{fig:post_compare} that the choice of prior mean doesn't have a large impact on the posterior distribution or MAP estimates. We attribute this to the fact that the RMSE is sufficiently small for both the ideal gas and PMF priors that the posterior distribution isn't significantly modified. However, it does influence the posterior predictive distribution as evidenced by Figure \ref{fig:postpredictive_compare}. Specifically, note  that there is uncertainty at low $r$ for the ideal gas prior, whereas this uncertainty vanishes for the PMF prior.

\subsection{Hyperparameter Selection}

The final step is to learn a set of LGP hyperparameters that provide accurate estimates of the target QoI. A standard approach to selecting hyperparameters is to maximize the model evidence \cite{rasmussen_gaussian_2006} or apply an expected improvement criterion based on an integrated acquisition function \cite{snoek_practical_2012}. Here we applied a brute force search based on minimizing the leave-one-out (LOO) error for 25,000 hyperparameter options randomly sampled over a prior range using the method of Sundararajan and coworkers \cite{sundararajan_predictive_2001} (Table \ref{tab:hyperparams}). This method gives relatively similar hyperparameter estimations for both an ideal gas and PMF GP prior. The prior range was selected based on the ($\lambda$-6) Mie parameter sensitivity analysis of Mick and coworkers \cite{mick_optimized_2015}.  

\begin{table}
\centering
\caption{Optimum hyperparameter values under the ideal gas and PMF prior computed from 25000 random samples over the reported test range.}
\begin{tabular}{| c | l | r | r |}
\hline
\textrm{Name}&
\textrm{Test Range}&
\textrm{Ideal Gas}&
\textrm{PMF}\\
\hline
$\ell_{\lambda}$  & 0.5-4      &  3.31   & 3.58\\
$\ell_{\sigma}$   & 0.01-0.05  &  0.046  & 0.048\\
$\ell_{\epsilon}$ & 0.001-0.01 &  0.0098 & 0.0093\\
$\alpha$          & 1E-4-0.1   &  0.094  & 0.095\\
$\sigma_{noise}$  & 1E-4-0.01  &  7.2E-4 & 8.3E-4\\
\hline
\end{tabular}
\label{tab:hyperparams}
\end{table}

A limitation of the brute force approach to hyperparameter selection is that we don't account for potential hyperparameter uncertainty in the LGP prediction. However, the self-consistency of our predictions with existing literature on liquid neon \cite{mick_optimized_2015, shanks_transferable_2022} suggests that this uncertainty is likely insignificant. Note that one could propagate hyperparameter uncertainty by performing Bayesian optimization over the hyperparameters, sampling the resultant hyperparameter posterior distribution, and propagating the samples through the posterior predictive estimation step. Finally, hyperparameter optimization for the LGP model is non-trivial since the LGP is an approximation to a non-stationary stochastic process \cite{heinonen_non-stationary_2016}. 

\begin{figure}
    \centering
    \includegraphics[width = 12cm]{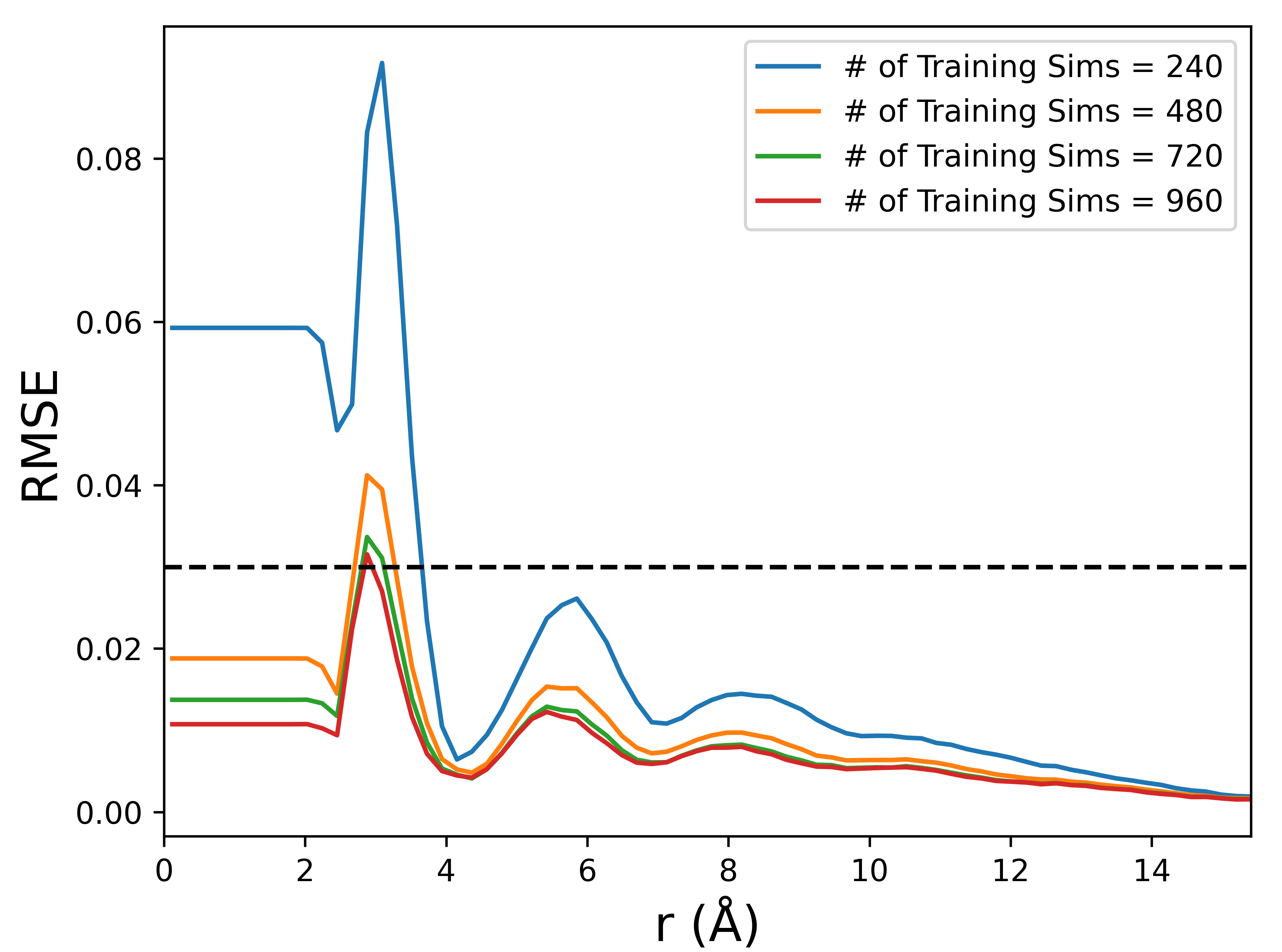}
    \caption{RMSE along $r$ for varying numbers of training simulations under an ideal gas prior.}
    \label{fig:rmsevssims}
\end{figure}

What if the previously described method fails to yield an accurate surrogate model? In this case, one can repeat the sequential sampling by adding more training simulations at each range to retrain the LGP until the RMSE is sufficiently small. As an example, Figure \ref{fig:rmsevssims} demonstrates that surrogate model accuracy improves as more samples are added at each range. Note that the accuracy of the surrogate will not improve beyond the statistical uncertainty of the underlying model.

A more rigorous, but non-trivial method for surrogate model training, is to use adaptive or on-the-fly learning, in which the uncertainty in the LGP prediction is used to decide whether or not a new simulation is needed in the training set. This approach has been used in prior work\cite{angelikopoulos_bayesian_2012, angelikopoulos_x-tmcmc_2015} but was found to be unnecessary for our purposes due to the efficiency and accuracy of the LGP with relatively few training samples.

\subsection{Using the LGP Surrogate Model for Parameter Sensitivity Analysis}

\begin{figure}
    \centering
    \includegraphics[width = 12cm]{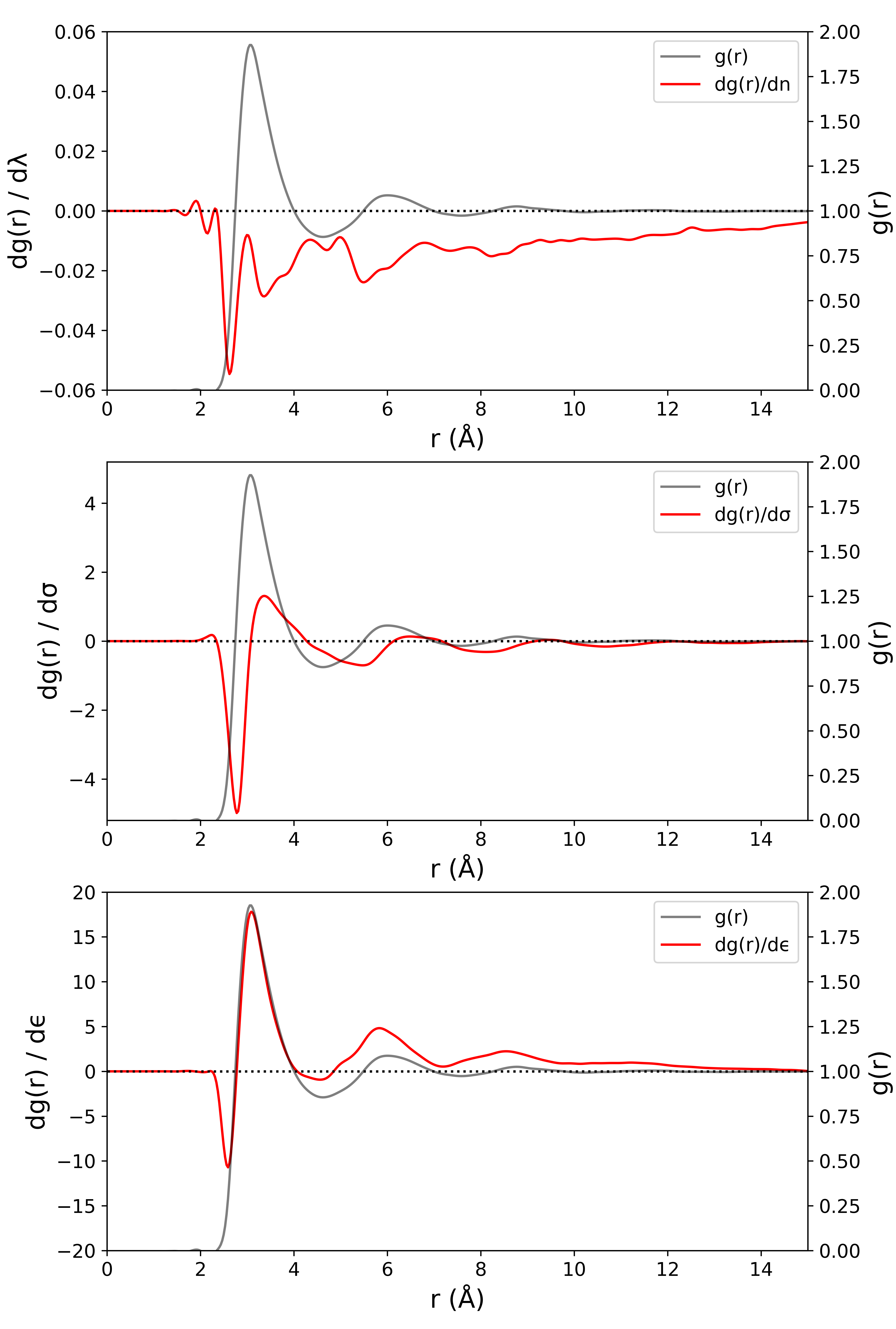}
    \caption{Derivatives of the local GP along the RDF calculated from eq \eqref{eq:derivs}.}
    \label{fig:derivs}
\end{figure}

Sensitivity of a QOI to a model parameter, $\theta_i$, can be quantified using the analytical derivative of the local GP surrogate model according to the following equation, 

\begin{equation}\label{eq:derivs}
    \frac{\partial \mathbb{E}[\textit{GP}_k(\boldsymbol{\theta}^*)]}{\partial \theta_i} = \bigg(\frac{ \theta_i - \theta_i^*}{\ell_{\theta_i}^2} \bigg) \mathbf{K}_{\boldsymbol{\theta}^*,\mathbf{\hat{X'}}} [\mathbf{K}_{\mathbf{\hat{X'}}, \mathbf{\hat{X'}}} + \sigma_{noise}^2 \mathbf{I}]^{-1}\mathbf{\hat{Y'}}_k
\end{equation}

\noindent where $k$ is the QoI index. The Gaussian process derivative is a quantitative measure of the influence of a perturbation in $\theta_i$ to the expectation of the observable $\mathbf{\hat{Y'}}_k$. Therefore, eq \eqref{eq:derivs} can approximate the impact of changing a molecular simulation parameter on the QOI (\textit{e.g.} how much does changing the effective particle size impact the RDF at any $r$). Figure \ref{fig:derivs} shows probabilistic local GP derivatives calculated for the ($\lambda$-6) Mie parameters. 

The repulsive exponent derivative exhibits a small magnitude and has a minimum at the RDF half maximum. This behavior suggests that increasing the repulsive exponent, which determines the "hardness" of the particles, steepens the slope of the first peak in the RDF. This result is intuitive considering that in a hard-particle model there is a discontinuous jump at the hard-particle radius (infinite slope) that progressively softens with the introduction of an exponential repulsive decay function. In the case of the collision diameter, zeros of the derivative occur at RDF peaks and troughs, while local extrema align with the half-maximum positions. Consequently, increasing the effective particle size shifts the RDF to the right while maintaining relatively constant peak heights. Regarding the dispersion energy, its derivative displays zeros at the half-maximum positions of the RDF and local extrema at peaks and troughs. This behavior indicates that an increase in the dispersion energy leads to an increased magnitude of the RDF peaks and greater liquid structuring. 

Derivatives of structure with respect to thermodynamic state variables ($T$, $P$, $\mu$, etc) can be computed with fluctuation theory. Let's now take as an example the $\epsilon$-derivative of the RDF in Ne. We find that an increase in the dispersion energy deepens the interatomic potential well, resulting in greater attraction and a more structured liquid. Noting that the reduced temperature, $T^*$, is inversely related to $\epsilon$ by,

\begin{equation}
    T^* = \frac{k_B T}{\epsilon}
\end{equation}

\noindent then the $g(r)$ derivative with respect to $\epsilon$ at constant $T$, is equal to the $g(r)$ derivative with respect to the reduced thermodynamic beta,

\begin{equation}
    \frac{\partial g(r)}{\partial \epsilon} = \frac{\partial g(r)}{\partial \beta^*}
\end{equation}

\noindent where $\beta^* = T^*/k_BT$. In summary, an increase in $\epsilon$ is equivalent to a decrease in temperature. It is therefore expected that the $\epsilon$ derivative and temperature derivative behave the same; specifically, a decrease in temperature should increase result in greater fluid structuring without significantly impacting peak positions. Unsurprisingly, this behavior is exactly what was observed in recent work that computed temperature derivatives of the O-O pair RDF in water using a fluctuation theory approach \cite{piskulich_temperature_2020}.

\section{The Standard Bayesian Framework}

For simplicity of notation, let $\theta = \{\lambda, \sigma, \epsilon, \sigma_n\}$ represent the model parameters and $\mathcal{Y} = S_d(Q)$ be the RDF observations. The nuisance parameter, $\sigma_n$, represents the width of the Gaussian likelihood and is considered a model parameter since nothing is known about this parameter \textit{a priori}. Calculating the posterior probability distribution with Bayesian inference then requires two components: (1) prescription of prior distributions on the model parameters, $p(\theta)$, and (2) evaluation of the RDF likelihood, $p(\mathcal{Y}|\theta)$. The prior distribution over the $(\lambda-6)$ Mie parameters is assumed to be a multivariate normal distribution,

\begin{equation}
    \theta \sim \mathcal{N} (\mu_\theta, \sigma^2_{\theta})
\end{equation}

\noindent where $\mu_\theta$ and $\sigma^2_\theta$ are the prior mean and variance of each $(\lambda-6)$ Mie parameter in $\theta$, respectively. A wide, multivariate normal distribution was selected because it is non-informative and conjugate to the Gaussian likelihood equation. The prior on the nuisance parameter is assumed to be log-normal,

\begin{equation}
    \log \sigma_{n} \sim \mathcal{N} (\mu_{\sigma_n}, \sigma_{\sigma_n}^2)
\end{equation}

\noindent where $\mu_{\sigma_n}$ and $\sigma_{\sigma_n}$ are the prior mean and variance of the nuisance parameter. The log-normal prior imposes the constraint that the nuisance parameter is non-negative, which is obviously true because a negative variance in the observed data is undefined. For reference, the prior parameters used in this study are summarized in Table \ref{tab:priors}.

\begin{table}
\centering
\caption{Prior parameters on the ($\lambda$-6) Mie model parameters.}
\begin{tabular}{| c | l | r r |}
\hline
\textrm{Parameter}&
\textrm{Distribution}&
\textrm{$\mu$}&
\textrm{$s$}\\
\hline
$\lambda$  &                   &  12.0    & 9\\
$\sigma$   & \text{Normal}     &  2.7     & 1.8\\
$\epsilon$ &                   &  0.112   & 0.225\\ \cline{1-4}
$\sigma_n$      & \text{Log-Normal} &  1       & 1\\
\hline
\end{tabular}
\label{tab:priors}
\end{table}

The likelihood function is assumed to be Gaussian according to the central limit theorem,

\begin{equation}
    p(\mathcal{Y}|\theta) \propto \frac{1}{\sigma_n^{n_{samples}}}\exp\bigg[-\frac{1}{2\sigma^2_{n}}\sum_i\ [{S}_{\theta_i}(Q_j) - S_d(Q_j)]^2\bigg]
\end{equation}

\noindent where ${S}_\theta(Q_i)$ is the molecular simulation predicted RDF and $j$ indexes over discrete points along the momentum vector. Bayes' theorem is then expressed as,

\begin{equation}\label{eq:inference}
    p(\theta|\mathcal{Y}) \propto p(\mathcal{Y}|\theta) p(\theta)
\end{equation}

\noindent where equivalence holds up to proportionality. This construction is acceptable since the resulting posterior distribution can be normalized \textit{post hoc} to find a valid probability distribution. 

\subsection{Markov Chain Monte Carlo}

To populate the Bayesian likelihood distribution, Markov Chain Monte Carlo (MCMC) samples over the model parameters $\theta = \{\lambda, \sigma, \epsilon, \sigma_n\}$  are passed to the surrogate model, evaluated, and compared to the experimental RDF. 960,000 MCMC samples were calculated using the emcee package \cite{foreman-mackey_emcee_2013} from 160 walkers (5000 samples/walker) with a 1000 sample burn-in per walker. The MCMC moves applied were differential evolution (DE) at a 0.8 ratio and DE Snooker at a 0.2 ratio, which is known to give good results for multimodal distributions. The acceptance ratio obtained from this sampling procedure was $\sim$0.27 and the autocorrelation between steps was 16 moves. 